# Changes in the Cloud Belts of Jupiter, 1630-1664, as reported in the 1665 *Astronomia Reformata* of Giovanni Battista Riccioli

Christopher M. Graney

Jefferson Community & Technical College

1000 Community College Drive

Louisville, Kentucky (USA) 40203

christopher.graney@kctcs.edu

www.jefferson.kctcs.edu/faculty/graney

A translation of a section from the 1665 *Astronomia Reformata* of G. B. Riccioli discussing the appearance of the disk of Jupiter during the years 1630-1664; changes in the Jovian cloud belts as recorded by a variety of observers are a major feature of Riccioli's discussion.

**Introduction**

The recent disappearance of one of Jupiter's cloud belts (Figure 1) has attracted significant attention from the scientific media (Phillips 2010). While this recent disappearance of a cloud belt is dramatic, changes in the clouds of Jupiter occur from time to time and have often been discussed in the scientific literature (for example, Cheng et. al. 2008, Satoh and Kawabata 1992). In fact, changes in the appearance of the Jovian clouds have been noted since the very beginning of telescopic astronomy. What follows is a translation of a discussion of reports on the appearance of Jupiter's disk from 1630 to 1664 that is contained in the *Astronomia Reformata* (1665) by the Italian Jesuit astronomer Giovanni Battista Riccioli (Figure 2).

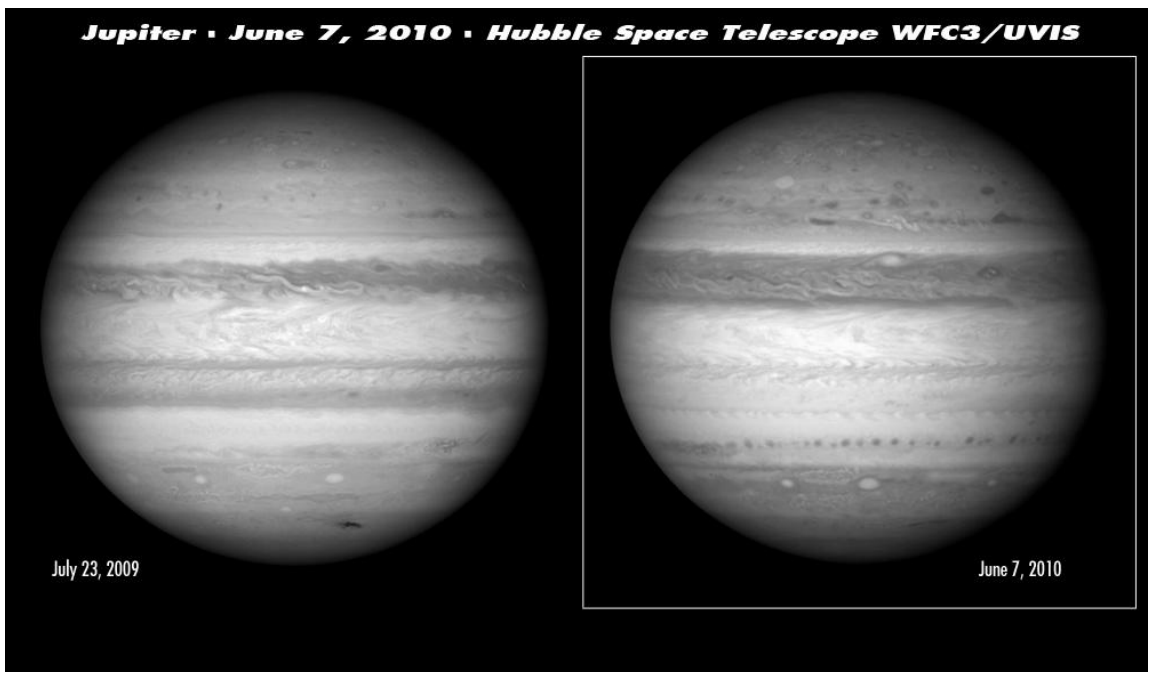

**Figure 1 – Images of Jupiter from the Hubble Space Telescope showing the 2010 disappearance of its South Equatorial Belt. Credit: NASA/ESA.**

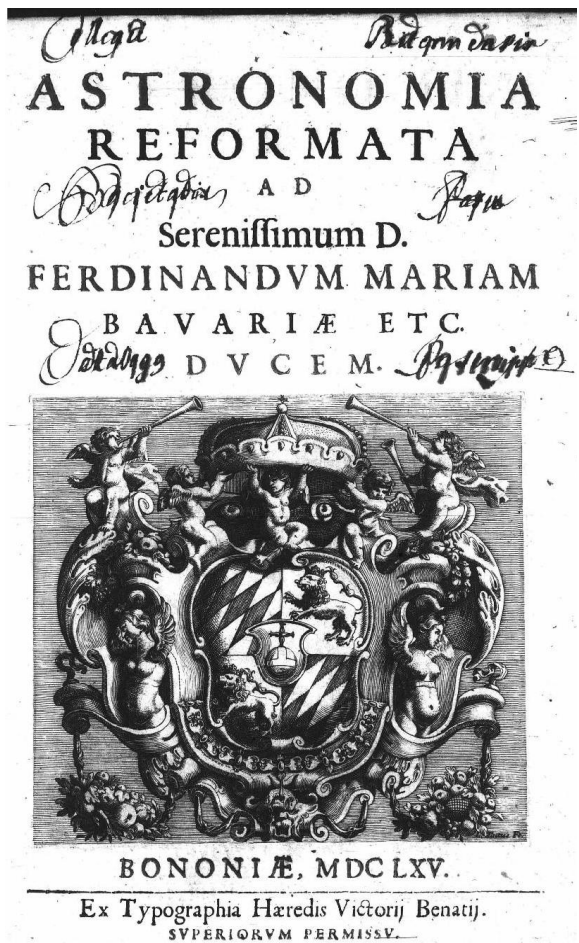

**Figure 2 – Title pages from Riccioli's *Astronomia Reformata*. According to the noted historian of science Edward Grant, unlike many of his colleagues who were "not scientists properly speaking but natural philosophers in the medieval sense using problems in Aristotle's *De caelo* and *Physics* as the vehicle for their discussions, Riccioli was a technical astronomer and scientist [Grant 1984, pg. 12]." Riccioli was very concerned with accuracy in measurements and observations, at one point recruiting nine of his Jesuit colleagues for the task of counting the swings of a pendulum over a twenty-four hour period, timed by transits of stars, in order to accurately calibrate it so he could time falling bodies – and then repeating the process to improve upon the results (Heilbron 2001, pp. 180-181). His reports on the appearance of Jupiter should be considered in light of this concern for accuracy.**

**Translation of *Astronomia Reformata* Chapter X, pages 368 through 370**

*Concerning the Figure of Jupiter, and its Bands[1], and their parallelism to the Ecliptic and Satellites.*

The atmosphere[2] which can be seen around Jupiter by means of a good telescope has been discussed by Father Anton Maria de Rheita [1645] in book 4, chapter 1, part 2 of *Radius Sidereomysticus*. The storminess of that atmosphere[3] was observed by Leander Bandtius, Abbot of Dunisburgh,[4] with his extraordinary telescope 1632 November 2, 21 days after opposition with the sun. He saw on Jupiter two small spots and two large spots, like hollows. One of the large spots was round. The other was oval, equaling one seventh of Jupiter's diameter at its longest.[5] This is according to a schema sent to me from Flanders; these features are seldom able to be seen, and then only by a telescope of exceptional quality and magnification.[6] Hevelius testifies to these phenomena in *Selenographia* [1647] pages 42 and 44. We shall relate more ordinary Jovian phenomena – the bands and the satellites – which are verified by observations.

The transverse bands that encircle Jupiter may be projections or depressions. Francesco Fontana may have observed them first from Naples, but the first recorded observations I know of were by Neapolitan Jesuit Fathers Giambattista Zuppi and Daniello Bartoli of Naples. Since then Father Francesco Maria Grimaldi and I have repeatedly and clearly observed them.

---

[1] The Latin word Riccioli uses is "fascia," which translates as "band" or "girdle", but also as "a streak of cloud".

[2] "Sphæra vaporosa"

[3] "De Iouis asperitate" – "asperitas" is roughness or harshness to the senses, so this might be best translated as the "unevenness" of Jupiter, but "asperitas" can be applied to rough weather, and "aspero" can be translated as "to make stormy". Since Riccioli is talking about an atmosphere, I have taken his words to refer to rough weather or storms.

[4] The *Collection of English Almanacs* (1790, page 46) contains a translation of this section and identifies Bandtius.

[5] The large oval suggests the Great Red Spot, whose discovery is typically attributed to Robert Hooke or Giovanni Domenico Cassini in the mid 1660's. See, for example, Bakich 2000, page 213.

[6] A more literal translation is "not able to be seen except by a rare Long-Looker" – "nec nisi raris Longispicijs discerni possunt".

Sometimes the bands appear three at a time. Sometimes they appear two at a time. Sometimes they appear singly. When single, sometimes they appear with slender boundaries on either side; other times they are just singular.

They may appear in the middle of the disk of Jupiter, on the underside, or on top. They are not always straight, but sometimes curved, and then they may curve up or down – manifest proof of the gyration (or libration around its center) of Jupiter. Information concerning how the bands are parallel to the ecliptic and to the path of the satellites is discussed separately below, following the section about the history of the band observations.

The bands are not always apparent, as is proven from the first schema of Giovanni Lipi of Monachiens.[7] Therefore it is not surprising, if Gassendi in book 3 of the *Institutio Astronomica* [1647] himself denies seeing them.

*Observations of Jupiter's Bands*

1630 May 17 – Father Niccolo Zucchi of Rome saw two bands of Jupiter as in II in Figure 3.

1633 – Fontana of Naples observed three bands, as in I in Figure 3 (as is contained in tract 6 of his observations). He recorded them again in 1643.

---

[7] "ex primo schemate Io. Lipij Monachiensis" – Giovanni Lipi is something of a guess, as I have not been able to identify this person – most likely another Jesuit, as searches for Monachiens seem to turn up references to a Jesuit College.

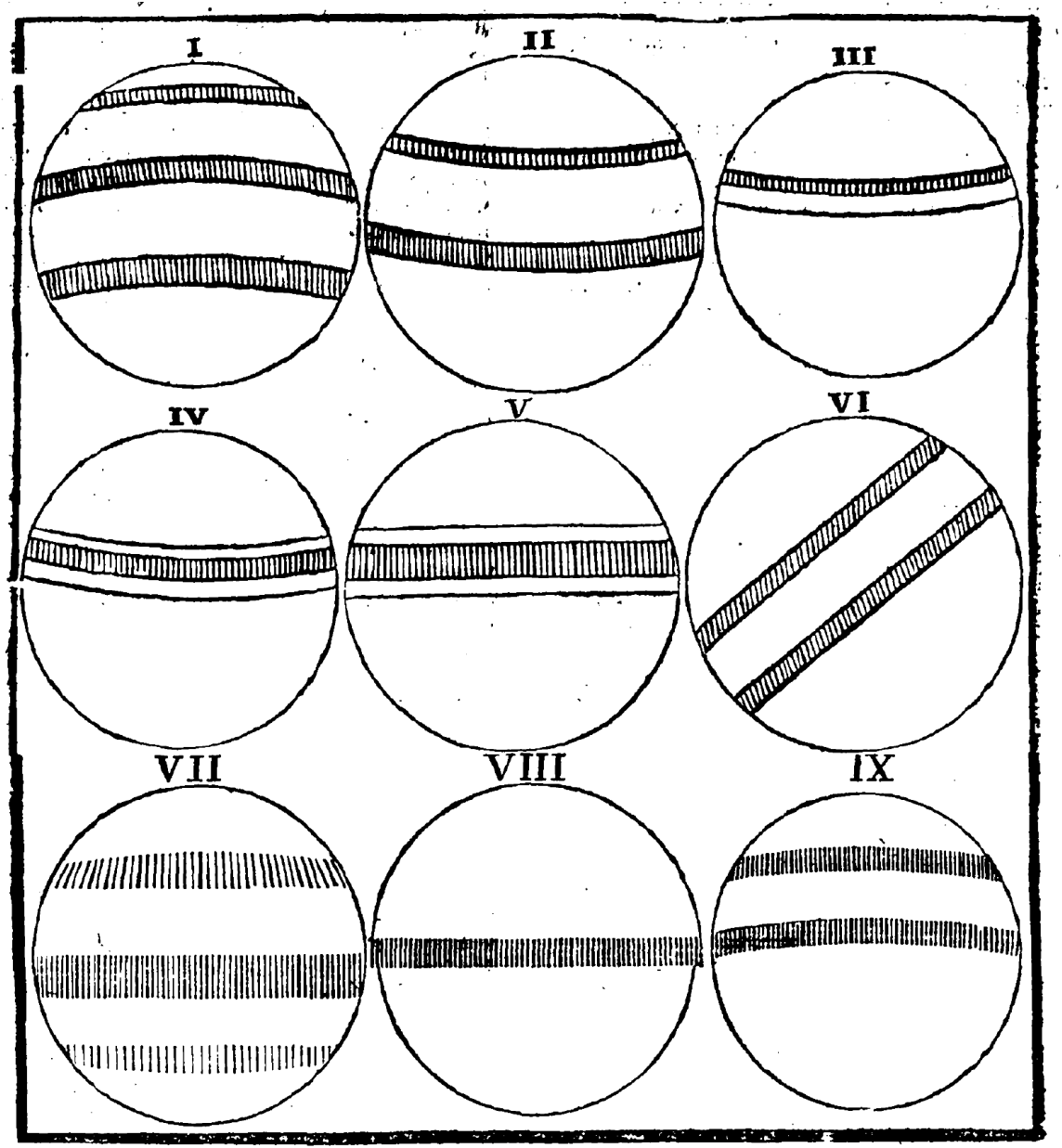

**Figure 3** – Figures from Riccioli's *Astronomia Reformata* used to illustrate the appearance of Jupiter. Some of the figures represent a specific observation. Others simply represent a general appearance that is referenced for multiple observations.

1634 – Jupiter had appeared daily with three bands. However, from the beginning of October all the way to October 13, around the time of opposition with the sun,[8] Grimaldi saw it to have two bands – one dark, the other bright by comparison to the rest of the Jovian disk, as in III in Figure 3. But after the opposition with the sun all the way to October 20, three curving bands could be seen, one dark, the others light in color as in IV. But from the beginning of November, all the way to December 22 they were always straight, as in V, particularly the southernmost one that ran through the middle of the disk.

1644 – Zuppi observed two bands of Jupiter, as in II in Figure 3 but much wider, on account of the excellence of the telescope, which showed the Jovian diameter measuring nearly half a foot.

1645 – On January 1 and 22 Fontana saw two Jovian bands, as in II in Figure 3, but by January 28 a third band had appeared near the edge, as in I.

1647 – On January 23 Grimaldi and I saw Jupiter in opposition with the sun encircled by two belts, as in II in Figure 3.

1648 – From February 24 to March 16 Grimaldi observed on Jupiter two straight bands, as in VI in Figure 3. This depicts the appearance of Jupiter on February 11, 2.5 hours past the setting of the sun. Jupiter was in the eastern quadrant somewhat apart from the meridian. The upper band was narrow and dark; the lower band was wider and less dark. The region between the bands was much brighter than the rest of the disk. Jupiter had been in opposition with the sun on February 22.

---

[8] Jupiter was not in oppostion with the sun in October of 1634. It was in opposition with the sun in October of 1631, so the 1634 date is probably erroneous and should be 1631. Typographical and other editing errors are common in Riccioli's work (Graney 2010).

1649 – On March 28 at the second hour of the night, bands appeared nearly in line with 4 satellites of Jupiter. But, on August 14, Grimaldi observed one band above the middle, in line with the satellites. The other bands were not clearly visible.

1650 – In the month of July, Grimaldi noticed two bands on the upper part of Jupiter, whose ends curved to the south as in IX in Figure 3.

1651 – On May 29, Grimaldi saw Jupiter to have two black bands which enclosed a large bright space as in II in Figure 3. On July 5, he saw three dark bands. The middle one was wider, darker, and just above the center of the disk. The southernmost was nearer to the edge somewhat like in VII.

1652 – Giovanni Battista Hodierna wrote to me of having seen Jupiter with one belt in June and July.

1653 – On July 21, Grimaldi saw three belts on Jupiter.

1654 – In January, Zuppi observed the lower half of the disk of Jupiter to have a smaller radius than the upper half as in VIII in Figure 3.

1656 – On August 2, Grimaldi noticed Jupiter to have two belts enclosing a brighter area as in VI in Figure 3. The lower belt was a little wider than the upper, and it was in line with the satellites. On October 21 and December 2, the bands looked like VII.

1657 – From October 17 to November 6, Grimaldi observed bands of Jupiter which resembled VII in Figure 3. But from November 12 to 19, Jupiter's appearance was changing. It was becoming less like VII and more like IX in that two black bands were enclosing a bright band. The third band near the southern limb was indistinct. Jupiter kept this appearance all the way to December 5. From then to the end of December it looked more like VII again but with a shadowy fourth little band on the southern edge.

1664 – On the first of July, Giuseppe Campani[9] observed Jupiter with the high-quality telescope he made himself. He saw four black bands and two bright ones. This is according to Doctor Giovanni Domenico Cassini. But I have not yet obtained the schema of them, so I cannot provide any details about them to the reader.

*Concerning the fact that the Bands, the Paths of the Satellites, and the Ecliptic are Parallel*

The bands of Jupiter, the path of its satellites, and the ecliptic are all parallel. Galileo [1610] in the *Sidereus Nuncius* and Simon Marius [1614] in *Mundus Jovialis* (part 2, section 6)[10] have recorded that the satellites and the ecliptic are parallel. Hevelius in *Selenographia* notes that the bands are likewise parallel with these. We have also noted this fact. Grimaldi or I often turned our attention toward the parallelism, which we confirmed by estimating a line through two satellites and comparing it to the bands and the path of all the satellites. It is not useful to fill up pages with these many observations.

**Conclusions**

Riccioli's discussion suggests that the Great Red Spot may have been observed decades earlier than history books tell us! More significantly, it suggests that changes in the appearance of Jupiter's cloud bands were noticed very shortly after the commencement of telescopic astronomy.

---

[9] A noted telescope maker. See King 2003, page 58.

[10] An English translation of this is available in Prickard 1916, page 404.

Graney, Changes in the Cloud Belts of Jupiter – page 9

Some of the changes Riccioli reports may be spurious. Riccioli is providing reports from a variety of astronomers, the qualities of whose instruments and observing skills may vary. Clearly the reports on the observations by Zuppi (1644, 1654) raise questions.

Nonetheless, there is no clear reason to question most of the reports. Those reports in which a single observer watches Jupiter over an extended period of time and notes changes in its appearance over that time should be particularly reliable. Many of the Grimaldi observations are examples of this, most notably those contained in the 1634 and 1657 reports.

Assuming the reports on the observations of Grimaldi and others are reliable, Jupiter's appearance was changing in the middle decades of the 17$^{th}$ century, much as its appearance changes today. Indeed, the reports in the *Astronomia Reformata* of Jupiter having two belts, and then one (1649), could almost be illustrated by Figure 1!

These reports on observations of Jupiter are more than just interesting items from astronomy's history – they are historical data. They show that the data record concerning Jupiter's appearance extends back nearly four centuries. Such data is potentially of value to modern astronomers. As libraries increasingly make historical material like Riccioli's *Astronomia Reformata* available via the internet, such data is becoming increasingly easy to find.

**Acknowledgements**

I thank Christina Graney for her invaluable assistance in translating Riccioli's writings.